\documentclass[aps,preprint,superscriptaddress]{revtex4-1}

\usepackage{amsmath}    
\usepackage{graphicx}   
\usepackage{color}      
\usepackage{subfigure}  
\usepackage{hyperref}   
\usepackage{todonotes}
\usepackage[latin1]{inputenc}

\begin{document}

\title{Exciton character in picene molecular solids}
\author{Friedrich Roth}
\author{Benjamin Mahns}
\author{Bernd B\"uchner}
\author{Martin Knupfer}
\affiliation{IFW Dresden, P.O. Box 270116, D-01171 Dresden, Germany}
\date{\today}

\begin{abstract}
We have studied the low-energy electronic excitations of solid picene at 20\,K using momentum dependent electron energy-loss spectroscopy. Our
results demonstrate the presence of five excitonic features below the transport energy gap of picene, which all are characterized by a
negligible dispersion. One of these excitons has not been observed in the optical absorption spectrum of picene molecules in solution and thus
is assigned to a (solid-state induced) charge transfer exciton. This conclusion is supported by the momentum dependent intensity variation of
this exciton which clearly signals a significant dipole forbidden contribution, in contrast to the other low energy excitations.
\end{abstract}

\maketitle

\section{Introduction}

Organic molecular solids consisting of aromatic hydrocarbons or related molecules have been in the focus of research for a number of reasons. In
particular their potential application in organic electronic devices has motivated many investigations in the past
\cite{Granstrom1998,Dodabalapur1995,Tsivgoulis1997,Kaji2009}. In addition, charge transfer, i.\,e. the addition or removal of charges to or from
molecules in organic solids is one route to modify and control their electronic properties. There are prominent examples, where charge transfer
is an essential ingredient in the control of the electronic properties of the respective material class. For instance, the formation of charge
transfer compounds is accompanied by fascinating, and sometimes unexpected physical properties. Well-known examples are the charge transfer
salts (e.\,g. TTF-TCNQ) with low dimensional electronic behavior and very rich phase diagrams \cite{Toyota2007}, or alkali doped fullerenes
\cite{Gunnarsson2004,Weaver1994,Gunnarsson1997}, which support superconductivity, or Mott insulating phases depending on the doping level. Also, charge
transfer induced by doping has been successfully applied to tune the electronic properties of organic semiconductor devices
\cite{pfeiffer1998,nollau2000,Gao2002}.

\par

Recently, the discovery of superconductivity with transition temperatures as high as 18\,K in alkali doped picene has added a new member to the
family of charge transfer compounds with fascinating physical properties \cite{Mitsuhashi2010}. This discovery asks for the investigation of the
physical properties of picene in the undoped and doped state in order to develop an understanding of the superconducting and normal state
properties, which also might help to tailor further superconductors on the basis of aromatic hydrocarbons.

\par

Picene (C$_{22}$H$_{14}$) is a molecule that consists of five benzene rings arranged in a zigzag like manner. In the condensed phase, picene
adopts a monoclinic crystal structure, with lattice constants $a$ = 8.48\,\AA, $b$ = 6.154\,\AA, $c$ = 13.515\,\AA, and $\beta$ = 90.46$^\circ$,
the space group is P2$_1$, and the unit cell contains two inequivalent molecules.\cite{De1985} The molecules arrange in a herringbone manner
which is typical for many aromatic molecular solids. Theoretical and experimental investigations recently have provided first insight into the
electronic properties of picene and its alkali metal doped relative. Undoped picene is characterized by a relatively large band gap in the
electronic spectrum and by four close lying conduction bands above the gap \cite{Roth2010}. These conduction bands are filled with electrons
upon potassium addition. A recent photoemission study has demonstrated the appearance of a new spectral structure in the gap of picene as a
function of K doping \cite{Okazaki2010}. The electronic structure of K doped picene has also been addressed recently using calculations
\cite{Giovannetti2010,Kim2010,Andres2010,Kosugi2009}. These indicate a filling of the conduction bands in K$_3$picene and the concomitant formation
of a Fermi surface.

\par

In this contribution we present a detailed experimental analysis of the electronic excitation spectrum of undoped solid picene using electron
energy-loss spectroscopy (EELS) in transmission. This technique enables the determination of the electronic excitation spectrum as a function of
momentum transfer and thus provides valuable insight into the dispersion and character of the excitations under scrutiny
\cite{Knupfer1999,Knupfer2000,Knupfer2002,Schuster2007,Kramberger2008}. Our studies are carried out at 20\,K and allow the identification of
five excitonic features below the transport gap of picene. These excitons are characterized by a negligible dispersion, which points towards
their rather localized nature. The momentum dependent intensity variation of the excitons provides interesting details on dipole allowed and
forbidden characters.

\section{Experimental}

\noindent Thin films of picene were prepared by thermal evaporation under high vacuum onto single crystalline KBr substrates kept at room
temperature with a deposition rate of 0.2\,nm/min. The resulting film thickness was about 100\,nm. These picene films were floated off in
destilled water, mounted onto standard electron microscopy grids and transferred into the EELS spectrometer. Prior to the EELS measurements the
films were characterized \textit{in-situ} using electron diffraction. All observed diffraction peaks were consistent with the crystal structure
of picene,\cite{De1985} while the diffraction spectra revealed a well pronounced texture. The films show a strong preference of crystallites with their $a,b$-plane
parallel to the film surface.\cite{Roth2010} Moreover, the width of the diffraction features in momentum space allows an estimation of the grain
size in our picene films of at least 70\,\AA, i.\,e. significantly larger than the lattice constants. In other words, a typical grain in our films
contains more than 800 molecules. Thus, our data represent the electronic excitations of crystalline solid picene.

\par

The EELS measurements were carried out at 20\,K using a 170\,keV spectrometer in combination with a He flow cryostat described
elsewhere.\cite{Fink1989} We note that at this high primary beam energy only singlet excitations are possible. The energy and momentum
resolution were chosen to be 85\,meV and 0.03\,\AA$^{-1}$, respectively. We have measured the loss function
Im[-1/$\epsilon(\textbf{q},\omega)$], which is proportional to the dynamic structure factor S($\textbf{q},\omega$), for momentum transfer
$\textbf{q}$ parallel to the film surface [$\epsilon(\textbf{q},\omega)$ is the dielectric function].

\section{Results and discussion}

\begin{figure}[ht]
 \includegraphics[width=0.6\textwidth]{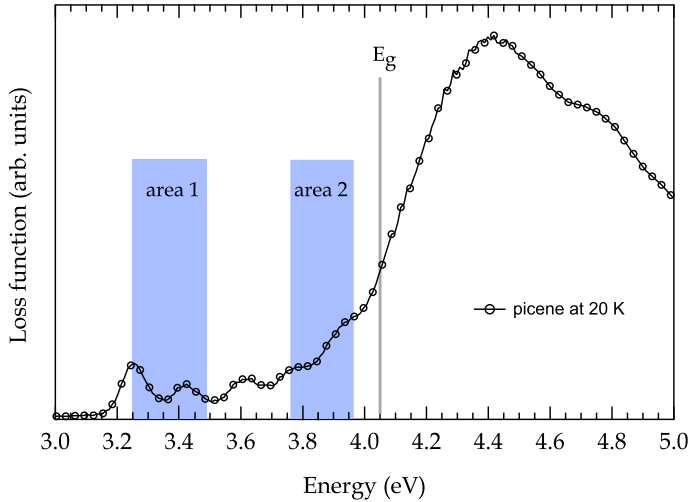}
 \caption{\label{fig1} Loss function of solid picene measured with a momentum transfer of $q$ = 0.1\,eV at 20\,K. The two emphasized areas indicate the energy ranges of former observed peaks in the optical absorption spectrum of picene in solution as reported in \cite{Gallivan1969,Ruzicka1936}. In addition, the transport energy gap, E$_g$ of solid picene is indicated.}
\end{figure}

In Figure \ref{fig1} we present the loss function of solid picene in a range between 3 - 5\,eV measured at 20\,K. We note that due to the
pronounced texture of our films,\cite{Roth2010} the data in Fig. \ref{fig1} predominantly represent excitations with a polarization vector
within the $a,b$ crystal plane. The loss spectrum as presented in Fig. \ref{fig1} is dominated by a broad peak at 4.4\,eV as well as a
pronounced fine structure right after the excitation onset. The overall structure of the loss function has been very well reproduced recently
using calculations based upon the GW self-energy approximation.\cite{Roth2010} In addition, in Fig. \ref{fig1} we can identify five well
separated features at 3.25, 3.41, 3.61, 3.77 and 3.93\,eV. Compared to previous EELS measurements of solid picene at room temperature
\cite{Roth2010} these low energy structures are significantly better resolved and well defined.

\par

In general, the lowest electronic excitations in organic molecular solids usually are excitons, i.\,e. bound electron-hole pairs
\cite{Pope1999,Silinsh1980,Knupfer2003,Lof1992,Hill2000}. This is one of the consequences of the weak van-der-Waals interaction between the molecules, which is
responsible for the molecular arrangement in the crystal. The decision criterion that has to be considered in order to analyse the excitonic
character and binding energy of an excitation is the energy of the excitation with respect to the so-called transport energy gap, which
represents the energy needed to create an unbound, independent electron-hole pair. This transport energy gap of picene has been estimated
previously to about 4.05\,eV \cite{Roth2010,Sato1987}. Consequently, the five excitation features in solid picene that are observed below 4\,eV as depicted in Fig.\,\ref{fig1} are excitons, and the exciton binding energy of the lowest lying exciton is as large as about 0.85\,eV.

\par

In addition, the electronic excitation spectrum of individual picene molecules has been studied in the past using optical absorption
measurements.\cite{Gallivan1969,Ruzicka1936} In Fig.\,\ref{fig1} we have highlighted the energy areas in which these optical data show
corresponding excitation structures.

Intriguingly, these studies reveal five excitations below 4\,eV, however no excitation feature has been observed so far in the energy window
around 3.61\,eV as highlighted in Fig.\,\ref{fig1}. In general, electronic excitations in solution and in the condensed phase can be observed at
different energies due to different screening effects related to the polarization of the surrounding. An assignment of the excitation at
3.61\,eV  to one of the features observed for picene molecules in solution would thus require a downshift of this excitation feature of about
200\,meV or more going to the condensed phase. However, we do not observe such a large shift for the lowest lying excitations, in their case the
difference between solution and solid state data is smaller than 40\,meV. As a consequence, an assignment of the excitation feature at 3.61\,eV
to a molecular electronic transition also seen in the optical absorption of picene molecules in solution would require an anisotropy of the
dielectric screening of more than a factor of five taking into account the different polarization of the excitations.\cite{Gallivan1969} Such a
large anisotropy however is very unlikely for a molecular crystal out of aromatic hydrocarbons. For instance, in the case of pentacene - a close
relative of picene - ellipsometry investigations of single crystals have revealed a maximal anisotropy of the dielectric constant along the
crystal axes of less than 1.8!\cite{Dressel2008} We therefore assign the exciton at 3.61\,eV to a solid-state induced electronic excitation, a
conclusion which is further supported by the momentum dependence of this excitation as discussed below.

 The most likely candidate for such an excitation is a charge transfer transition,
where in the final state the electron and hole sit on adjacent picene molecules. This is reminiscent of the low energy excitations of pentacene,
a close aromatic relative of picene, where also such charge transfer excitations have been discussed.\cite{Schuster2007,Tiago2003,Hummer2005}
Within a simple point charge approach,\cite{Pope1999} one can estimate the binding energy $E_B$ of such a charge transfer exciton using

\begin{align*}
 E_B \sim \frac{1}{4\pi\epsilon_0\epsilon_r} \frac{e^2}{\langle r \rangle},
\end{align*}

where $\langle r \rangle$ denotes the mean distance of the two adjacent molecules which participate in the charge transfer excitation, and
$\epsilon_r$ is the static dielectric constant ($\epsilon_r$ $\sim$ 4 for picene \cite{Roth2010}). The distance of the two adjacent molecules in
the $a,b$ plane of solid picene is about 5.3\,\AA, which leads to an estimate for $E_B$ of about 0.7\,eV. In other words, in picene it is
reasonable to assume charge transfer excitons at similar energies as intra-molecular excitons, also called Frenkel excitons. Moreover, the
presence of both types of excitons at similar energies can lead to a sizable interaction of these excitation species and result in excitons with
a mixed character, a situation that has been discussed in the past also for other organic molecular solids. \cite{Hoffmann2007,Knupfer2004,Tokura1983}

\par

\begin{figure}[ht]
 \includegraphics[width=0.6\textwidth]{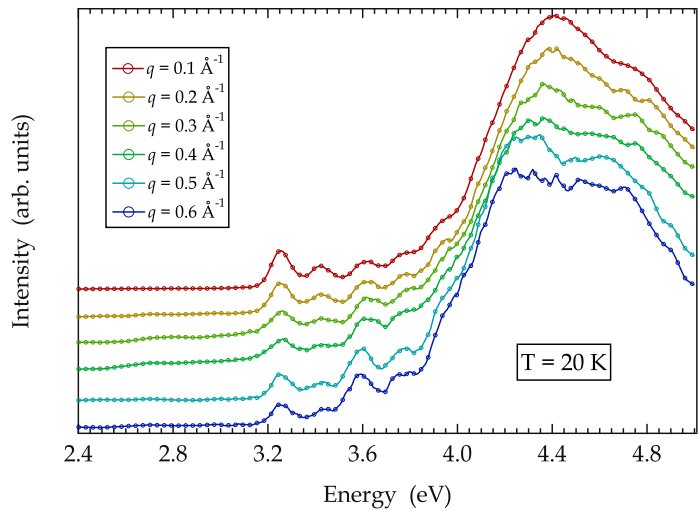}
 \caption{\label{fig2}Momentum dependence of the EELS spectra of solid picene ($q$ is increasing from top to bottom). The measurements were carried out at 20\,K.
}
\end{figure}

In order to obtain a more detailed picture of the excitons in picene we have measured the loss function with increasing momentum transfer $q$. As shown in Figure \ref{fig2}, all identified excitons in picene with the exception of the feature at 3.61\,eV do not change in energy within a
momentum range up to 0.6\,\AA, which covers almost the entire first Brillouin zone parallel to the $a,b$ crystal directions. Consequently, in
the framework of an exciton band structure description this yields a vanishing group velocity ($\sim \frac{\partial E(k)}{\partial k}$) for
these excitations, i.\,e. they can be regarded as rather localized. For the excitation at 3.61\,eV a very small negative dispersion of about 30\,meV
can be seen in Fig.\,\ref{fig2}. In consideration of the assignment of this excitation to a charge transfer exciton (see above), this finite
dispersion would corroborate the different character of this excitation, we however emphasize that the size of the dispersion is close to the
experimental resolution.

\begin{figure}[t]
  \includegraphics[width=0.49\textwidth]{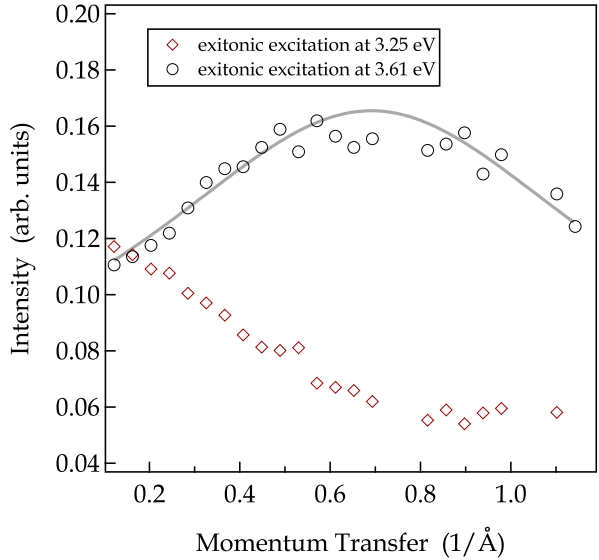}
 \caption{\label{fig3}Momentum dependence of the spectral weight of the two excitonic excitations at 3.25\,eV (red diamonds) and 3.61\,eV (black circles) as observed
 in the loss function of solid picene. The data around 0.75\,\AA$^{-1}$ and 1.05\,\AA$^{-1}$ could not be determined
 accurately enough because of considerably enhanced multiple scattering in this region due to (100) and (010) Bragg reflections. The data are normalized
 to the $q$ dependent intensity variation at 10\,eV excitation energy in order to take into account the
 overall momentum dependence of the scattering cross section.}
\end{figure}

Interestingly, we can also identify a substantial intensity variation, especially for the three excitons lowest in energy. Moreover, while the
intensity of the first two excitons at 3.25 and 3.41\,eV decreases with increasing momentum transfer $q$, the opposite is the case for the
exciton observed at 3.61\,eV. In case of localized excitations, i.\,e. those with negligible dispersion, their character can be analyzed in terms
of a multipole expansion, whereas upon increasing momentum transfer dipole (or optically) allowed excitations loose intensity and dipole
forbidden excitions (e.\,g. quadrupole transitions) will increase in intensity.\cite{Knupfer1999,Knupfer1999_2,Knupfer2000,Haverkort2007} Thus,
the two singlet excitons with lowest excitation energy in solid picene are of predominant dipole character, while the following exciton at
3.61\,eV is characterized by a significant dipole forbidden contribution.  Furthermore, the momentum value $q_{max}$ where a dipole forbidden
excitation reaches its intensity maximum can be used to estimate the mean radius of the wave function of this excitation: $\langle r \rangle$
$\sim$ 2/$q_{max}$.\cite{Knupfer1999_2,Haverkort2007} In Figure \ref{fig3} we present a comparison of the intensity variation of the excitons at
3.25 and 3.61\,eV. Again, the decreasing intensity for the lowest lying exciton clearly signals its predominant dipole allowed character in good
agreement to the fact that this feature was also observed in optical absorption measurements of picene molecules in
solution.\cite{Gallivan1969,Ruzicka1936} In contrast, the intensity of the 3.61\,eV exciton reaches its intensity maximum at finite momentum (about 0.7\,\AA$^{-1}$) as would be theoretically expected for
e.\,g. a quadrupole excitation.\cite{Haverkort2007} This underlines a significant dipole forbidden part and now gives a (very rough) estimate of
the radius of this exciton of about 3\,\AA. Here, one should keep in mind that our data also suggest a mixed character of the excitons in solid picene (see discussion above) which limits a
quantitative analysis of the exciton extension. Nevertheless, the observed momentum maximum is in reasonable agreement to our interpretation of
the exciton at 3.61\,eV having charge transfer character.

\section{summary}

To summarize, our electron energy-loss spectroscopy studies at low temperature and as a function of momentum transfer have enabled to elucidate
the low energy singlet excitations in solid picene. The electronic excitation spectrum consists of five excitons below the transport energy gap
of picene. While four of those have also been observed in optical absorption measurements of individual picene molecules in solution, one of
these excitons (at 3.61\,eV excitation energy) is only seen in the condensed phase, and we attribute it to a charge transfer excitation involving
neighboring molecules in the crystal. Moreover, this solid state specific exciton is also distinguished by a significant dipole forbidden
character as revealed by our momentum dependent investigations. The fact that the lowest exciton has a binding energy of about 0.85 eV also
indicates that electronic correlation effects play a role in the electronic properties of picene,\cite{Lof1992} since in a simple approach this
binding energy can be taken as a measure for the electronic correlation energy $U$ and a value of $U$ $\sim$ 0.85\,eV is large than the
conduction band width.\cite{Roth2010}

\begin{acknowledgments}
We thank R. Sch\"onfelder, R. H\"ubel and S. Leger for technical assistance. This work has been supported by the Deutsche Forschungsgemeinschaft
(grant numbers KN393/13 and KN393/14). We are grateful to Matteo Gatti, Pierluigi Cudazzo, and Angel Rubio for fruitful discussions.
\end{acknowledgments}

\bibliography{Pic_exciton}
\bibliographystyle{apsrev4-1}
\end{document}